\documentclass[journal=jpclcd,manuscript=letter]{achemso}
\setkeys{acs}{articletitle = true}
\usepackage{achemso}
\usepackage{amssymb,amsfonts,amsmath}
\usepackage{times}
\usepackage{graphicx}
\usepackage{color}
\usepackage{orcidlink}
\usepackage{hyperref} \hypersetup{colorlinks=true,citecolor=blue,linkcolor=blue,urlcolor=blue} 

\def\BUFFALO{\footnotesize Department of Chemistry, State University of New York at Buffalo, Buffalo, NY, 14260, USA}
\def\BERKELEY{\footnotesize Department of Earth and Planetary Science, University of California, Berkeley, CA, 94720, USA}
\def\BERKELEYALT{\footnotesize Department of Astronomy, University of California, Berkeley, CA, 94720, USA}

\author{Xiaoyu Wang\orcidlink{0000-0001-7549-6010}}\affiliation{\BUFFALO}
\author{Nisha Geng}\affiliation{\BUFFALO}
\author{Kyla de Villa\orcidlink{0000-0001-6718-5095}}\affiliation{\BERKELEY}
\author{Burkhard Militzer\orcidlink{0000-0002-7092-5629}}\affiliation{\BERKELEY}\affiliation{\BERKELEYALT}
\author{Eva~Zurek\orcidlink{0000-0003-0738-867X}} \email{ezurek@buffalo.edu}\affiliation{\BUFFALO}

\title{Superconductivity in Dilute Hydrides of Ammonia under Pressure}

\begin{document}

\begin{tocentry}
\includegraphics[width=2in]{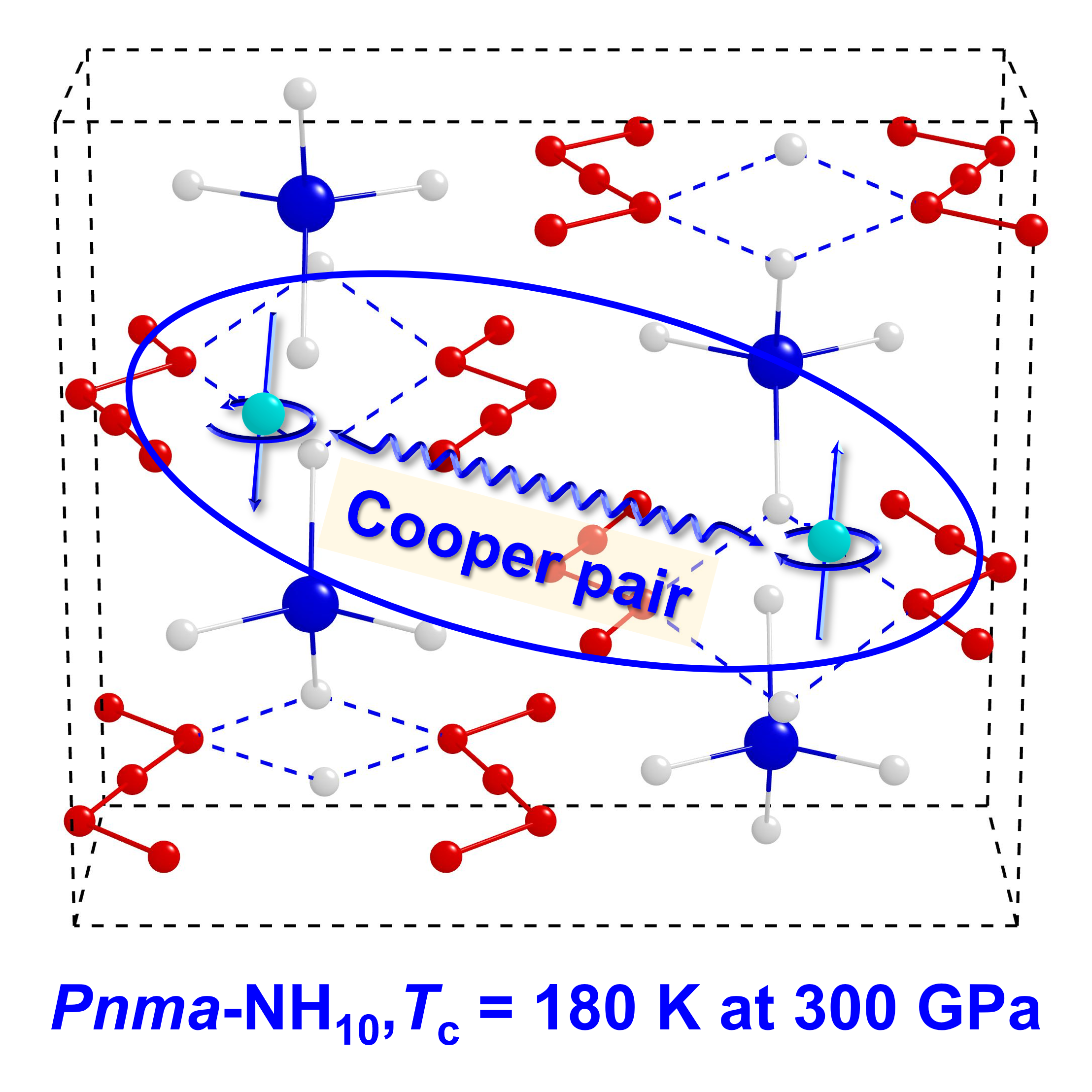}
\end{tocentry}

\begin{abstract}
In the last decade, there has been great progress in predicting and synthesizing polyhydrides that exhibit  superconductivity when squeezed. Dopants allow these compounds to become metals at pressures lower than those required to metallize elemental hydrogen. Here, we show that by combining the fundamental planetary building blocks of molecular hydrogen and ammonia,  conventional superconducting compounds can be formed at high pressure. Through extensive theoretical calculations we predict metallic metastable structures with NH$_n$ ($n=10,11,24$) stoichiometries that are based on NH$_4^+$ superalkali cations and complex hydrogenic lattices. The hydrogen atoms in the molecular cation contribute to the superconducting mechanism, and the estimated superconducting critical temperatures, $T_\text{c}$s, are comparable to the highest values computed for the alkali metal polyhydrides. The largest calculated (isotropic Eliashberg) $T_\text{c}$ is 179~K for $Pnma$-NH$_{10}$ at 300~GPa. Our results suggest that other molecular cations can be mixed with hydrogen under pressure yielding superconducting compounds.
\end{abstract}
\clearpage




The long-hypothesized phase of metallic hydrogen~\cite{Wigner1935_JChemPhys} tantalizes with exotic properties including high-temperature superconductivity~\cite{Ashcroft1968_PhysRevLett} and superfluidity~\cite{Babaev}. Though numerous groups have attempted to create this sought-after phase, reports of metallicity in statically compressed samples remain controversial~\cite{Gregoryanz2020_MatterRadiatExtremes}, and superconductivity has not been measured. However, a route towards lowering the metallization pressure, based upon the addition of an impurity element that can ``chemically precompress'' the hydrogen lattice~\cite{Carlsson1983_PhysRevLett,Ashcroft2004_PhysRevLett} has set the stage for the prediction and discovery of a series of hydrogen-based conventional superconductors at pressures found in Earth's core. Most of the studied compounds are covalent metals that contain hydrogen in combination with an electropositive metal or a $p$-block element~\cite{Zurek:2021k}. In this paper, we theoretically investigate the structure and superconducting properties of NH$_n$ ($n=7-24$) polyhydrides that are based on hydrogen and ammonia molecules, which are building blocks of giant planets. 

\emph{Ab initio} crystal structure prediction (CSP) studies have proposed a plethora of stable or low-lying metastable phases formed from doping hydrogen by a Group-I element under pressure~\cite{Zurek2017}. Numerous exotic structural motifs such as stretched H$_2^{\delta -}$ molecules (\textit{e.g.}\ in LiH$_6$ \cite{Zurek:2009c}), hydridic H$^-$ ions (\textit{e.g.}\ in NaH$_9$ \cite{Zurek:2011d}), linear or ever-so-slightly bent H$_3^-$ units (\textit{e.g.}\ in KH$_5$~\cite{Hooper2012_JPhysChemC}, RbH$_5$~\cite{Hooper2012_ChemAEurJ} and CsH$_3$~\cite{Zurek:2012g}) and (H$_3^-$)$_\infty$ polymeric chains (\textit{e.g.}\ in RbH$_6$~\cite{Hooper2012_ChemAEurJ}) are present in the CSP-found phases. The trends in their computed superconducting critical temperatures, $T_c$s, are related to their structural features, with compounds containing H$^-$ and H$_3^-$ units typically being poor candidates for superconductivity, those accommodating H$_2^{\delta -}$ units exhibiting intermediate $T_c$s, and phases with non-molecular hydrogenic motifs showing the most promise~\cite{Zurek2019_JChemPhys}.  The computed $T_c$s of polyhydrides of the alkaline earths are typically higher than those of the alkali metals~\cite{Zurek:2021k}. Nonetheless, various  superconducting Group-I high hydrides have been predicted including: $R\bar{3}m$-LiH$_6$ ($T_\text{c}=$~38~K at 150~GPa and 82~K at 300~GPa~\cite{Xie2014_ActaCrystC}, or 130-160~K at 300~GPa~\cite{Shipley2021_PhysRevB}), $I422$-LiH$_8$ (31~K at 100~GPa~\cite{Xie2014_ActaCrystC}), $C2/c$-KH$_6$ ($\sim$70~K at 166~GPa~\cite{Zhou2012_PhysRevB}), $Immm$-KH$_{10}$ (140~K at 150~GPa)~\cite{Semenok2020_CurrOpin}, and $C2/m$-RbH$_{12}$ (133~K at 150~GPa)~\cite{Semenok2020_CurrOpin}, $Pm\bar{3}m$-NaH$_6$ (260~K at 100~GPa) and $P6_3/mmc$-NaH$_9$ (252~K at 500~GPa)~\cite{Shipley2021_PhysRevB}.

Encouragingly, lithium polyhydrides have been synthesized under pressure (though their structure has not been determined)~\cite{Pepin2015_PNAS}, both NaH$_7$ and NaH$_3$ have been made below a megabar~\cite{Struzhkin2016_NatCommun,Marqueno2024_FrontChem}, and recently the synthesis of cesium and rubidium polyhydrides has been reported.\cite{Zhou2024_AEM} Density functional theory (DFT) calculations predict the stabilization pressure of these systems to decrease with decreasing ionization potential (IP) of the metal, so whereas the enthalpy of formation, $\Delta H_\text{F}$, of the lithium polyhydrides becomes negative around 120~GPa~\cite{Chen2017_InorgChem}, the cesium polyhydrides are stable by a mere 2~GPa~\cite{Zurek:2012g}. Quasi-spherical molecular complexes, such as Li(NH$_3$)$_4$, possessing frontier molecular orbitals with the same symmetry as those of the Group-I elements, but extending outside the component atoms, can be thought of as superalkali atoms~\cite{Zurek2009_AngewChemIntEd}.  We therefore wondered if other molecular species, with superatom characteristics resembling the Group-I elements, could be combined with hydrogen under pressure to access polyhydrides with unique chemical compositions? And, would the hydrogen atoms comprising the molecular complex help enhance their superconducting properties?

The NH$_4$ radical is a Rydberg molecule, whose singly occupied molecular orbital (SOMO) is extremely diffuse. The MOs of this tetrahedral molecule are reminiscent of an alkali metal atom's, with the nearly spherical $1a_1$ SOMO resembling the $s$-orbital, and the higher lying unoccupied triply degenerate $1t_2$ orbitals being of $p$-type symmetry (Figure~S2). Comparison of the radial distribution functions of these orbitals shows that the profile computed for NH$_4$ resembles most closely that of K or Rb (Figure~S3). The experimentally measured IP of NH$_4$, 4.62~eV~\cite{Fuke1994_ChemPhysLett},  falls in-between the IPs of Na and K (5.14 and 4.34~eV, respectively~\cite{lide1992ionization}), and is somewhat higher than the IP of Rb (4.18~eV~\cite{lide1992ionization}). Thus, based upon its IP and the radial extent of its SOMO, NH$_4$ is analogous to Rb or K. 

In the gas phase NH$_4$ is a short-lived metastable species that readily decomposes into NH$_3$ and a hydrogen radical. In condensed phases and above pressures of $\sim$90~GPa computations predict that NH$_3$ self-ionizes into phases containing NH$_4^+$ and NH$_2^-$ ions, driven by a volume decrease~\cite{Pickard2008_NatMater}.  Moreover, CSP searches predict that the ammonium cation is found in a number of stable high pressure phases including  NH$_4^+\cdots$OH$^-$ by 5~GPa~\cite{Griffiths2012_PhysRevB} and (NH$_4^+$)$_2\cdots$O$^{2-}$ by 65~GPa~\cite{Robinson:2017a}, and in NH$_3$-HF mixtures of varying composition~\cite{Conway:2021}. Support for the existence of NH$_4^+$ in a subset of these systems has subsequently been obtained in high-pressure experiments~\cite{Palasyuk:2014,Liu:2017}. Especially relevant for the work presented here, CSP studies on NH$_n$ ($n=1-9$) found that NH$_7$ has the most negative $\Delta H_\text{F}$ (from H$_2$ and NH$_3$) between 50-200~GPa~\cite{Song2019_JPhysChemLett}. Below 60~GPa the most stable NH$_7$ phase (of $R\bar{3}m$ symmetry) contained NH$_3$-H-NH$_3$, as well as H$_2$ and H$^-$ units, and at higher pressures a $P4_12_12$ structure, which remained insulating to at least 100~GPa, with tetrahedral NH$_4^+$, H$_2$ and H$^-$ units was preferred instead. 

Given the relative ease of formation of the ammonium ion under pressure and its similarity with that of an alkali-metal cation, we hypothesized that superconducting ammonium polyhydrides, with NH$_4$H$_n$ ($n\ge5$) stoichiometries, might be (meta)stable under pressure. Therefore, CSP searches were performed (using the \textsc{XtalOpt} evolutionary algorithm version 12~\cite{zurek:2011a,zurek:2020i}) for $n=3,5-12, 16, 20$ at 100, 200 and 300~GPa with the projector-augmented wave~\cite{Blochl1994_PAW} DFT~\cite{Kresse1993_VASP} framework (see Section S1 for further details). The many predicted low enthalpy structures were filtered, and only those that were metallic, high-symmetry, as well as frozen-phonon-dynamically and either meta- or thermodynamically stable were considered for further analysis and $T_\text{c}$ calculations. Thermodynamically stable structures lay on the (0~K) convex hull generated using NH$_3$ and H$_2$ as the endpoints, and  metastable ones were defined as those within 50~meV/atom of the hull (Figure~S1E-F).

\begin{figure*}[!ht]
    \centering   
    \includegraphics[width=1.0\textwidth]{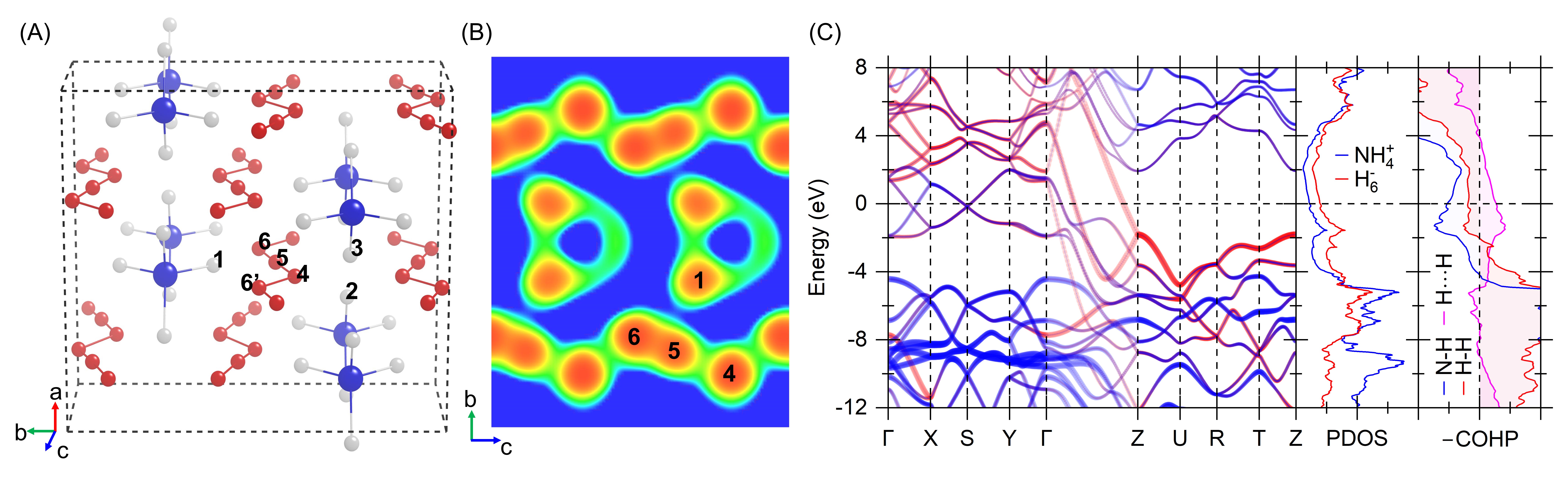}
    \caption{$Pnma$-NH$_{10}$ at 300~GPa: (A) Unit cell. N atoms are blue, H atoms in NH$_4^+$ are white and the H atoms comprising the 1D chains are red. (B) Plot of the Electron Localization Function (ELF) using an isosurface value of 0.5, with the contours colored from 0.5 (blue) to 1.0 (red). (C) Atom projected band structure and density of states showing contributions from atoms within the NH$_4^+$ units (blue) and atoms within the 1D chains (red). $-$COHP averaged over N-H bonds within  NH$_4^+$ units (blue), H-H bonds within the 1-D chain (red), and between the H atoms within NH$_4^+$ and those in the chain (pink).}
    \label{fig:pnma}
\end{figure*}

An intriguing phase that emerged in our CSP searches, $Pnma$-NH$_{10}$  (Figure~\ref{fig:pnma}A), was dynamically stable above 280~GPa (Figure~S4A). Though it was unstable with respect to decomposition into H$_2$ and NH$_3$ ($\Delta H_\text{F} \sim$23~meV/atom at 300~GPa within the PBE functional~\cite{Perdew1996_PBE}), inclusion of dispersion~\cite{grimme2010consistent} lowered the magnitude of the instability ($\Delta H_\text{F} \sim$7.5~meV/atom), and the zero-point energy (ZPE) corrections were further stabilizing, resulting in an overall $\Delta H_\text{F}$ of $\sim$6.5~meV/atom. This phase can be described as sheets of NH$_4^+$ molecules that lie in the $bc$ plane and are separated by one-dimensional (1D) zig-zag hydrogenic chains that run along the $c$-axis.  H$_2$ units (labelled as H(5) and H(6)) whose bond lengths are slightly stretched relative to those in the most stable H$_2$ phases at 1~atm ($P6_3/m$) or 300~GPa ($Cmca$-H$_2$), and whose Bader charges assign them as being nearly neutral, and hydridic hydrogens (H(4)), with a Bader charge of -0.15 comprise these chains.

The crystal orbital Hamilton population integrated to the Fermi level, $E_\text{F}$, (-ICOHP)~\cite{Dronskowski1993_COHP,Maintz2016_LOBSTER}, which gives a measure of the covalent bond strength, was smaller for the dihydrogen units within $Pnma$-NH$_{10}$ than for the H-H bonds within $P6_3/m$ and $Cmca$-H$_2$ (Table \ref{tab:bond-lenght-icohp}). Visualization of the electron localization function (ELF) (Figure~\ref{fig:pnma}B) revealed an even weaker covalent H-H$\cdots$H$^-$ interaction (with the minimum ELF values between the atom pairs being apparently greater than 0.5, the value characteristic of a free electron gas). Nonetheless, these three hydrogens can be thought to comprise an asymmetric H$_3^-$ molecule, whose bond lengths are more equalized than in the gas phase local minimum, predicted to possess one short (0.75~\AA{}) and one long (2.84~\AA{}) H-H distance via \emph{ab initio} calculations \cite{Ayouz:2010a}. The gas phase species possesses a double well minimum, with a symmetric H$_3^-$ transition state, corresponding to the simplest example of a three centered four electron bond. Previous CSP calculations revealed such H$_3^-$ motifs, which were symmetrized under pressure, within predicted RbH$_3$, RbH$_5$~\cite{Hooper2012_ChemAEurJ} and KH$_5$~\cite{Hooper2012_JPhysChemC} phases. Within $Pnma$-NH$_{10}$ the nearest neighbor intermolecular H$_3^-$ distances are only somewhat longer than the longest intramolecular distance (\emph{cf.}\ H(4)-H(6$^\prime$) \emph{vs.}\ H(4)-H(5)), suggesting that it may be more appropriate to view these hydrogenic motifs as a series of parallel $^1_\infty$[H$_2$$\cdots$H$^-$] chains, resembling those predicted in $Imma$-RbH$_6$~\cite{Hooper2012_ChemAEurJ} and  $R\bar{3}m$-SrH$_6$~\cite{Hooper2014_JPhysChemC} phases near 250~GPa.

\begin{table}[!htpb]
    \centering
    \setlength{\tabcolsep}{12pt}
    \begin{tabular}{lcc}
    \hline\hline
    bond type &  distance (\AA{}) & $-$ICOHP (eV/bond) \\
    \hline
    $P6_3/m$-H$_2$  (1~atm)& & \\
    \hline
    H$\textsubscript{mol}$-H$\textsubscript{mol}$  & 0.750  &  5.51 \\
    H$\textsubscript{mol}$-H$\textsubscript{mol}$  & 0.751  &  5.51 \\
    \hline
    $Cmca$-H$_{2}$ (300~GPa)  &  &  \\
     \hline
    H$\textsubscript{mol}$-H$\textsubscript{mol}$  & 0.756  &  6.15 \\
    H$\textsubscript{mol}$-H$\textsubscript{mol}$  & 0.775  &  5.78 \\
    H$\textsubscript{mol}$-H$\textsubscript{mol}$  & 1.100  &  1.45 \\
    H$\textsubscript{mol}$-H$\textsubscript{mol}$  & 1.131  &  1.23 \\
    H$\textsubscript{mol}$-H$\textsubscript{mol}$  & 1.146  &  1.18 \\    
    \hline
    $Pnma$-NH$_{10}$ (300~GPa)  &  &  \\
    \hline
    H5$\textsubscript{chain}$-H6$\textsubscript{chain}$    &  0.812  &  4.71 \\
    H4$\textsubscript{chain}$-H5$\textsubscript{chain}$    &  0.928  &  3.10 \\
    H4$\textsubscript{chain}$-H6$^{\prime}\textsubscript{chain}$    &  0.965  &  2.90 \\
    H1$\textsubscript{ammon}$-H5$\textsubscript{chain}$    &  1.047  &  1.92 \\
    H3$\textsubscript{ammon}$-H4$\textsubscript{chain}$   &  1.136  & 1.38 \\
    H2$\textsubscript{ammon}$-H4$\textsubscript{chain}$    &  1.173  &  1.19 \\
    \hline
    $Cmc2_1$-NH$_{11}$ (300~GPa)   &  &  \\
    \hline
    H5$\textsubscript{chain}$-H6$\textsubscript{chain}$  &  0.785  & 5.63 \\
    H3$\textsubscript{chain}$-H4$\textsubscript{chain}$  &  0.935  & 3.27 \\
    H4$\textsubscript{chain}$-H5$\textsubscript{chain}$  &  0.955  & 3.04 \\
    H1$\textsubscript{ammon}$-H3$\textsubscript{chain}$  &  1.067  &  1.51 \\
    H4$\textsubscript{chain}$-H6$^{\prime}\textsubscript{chain}$  &  1.100  &  1.60 \\
    H1$\textsubscript{ammon}$-H3$^{\prime}\textsubscript{chain}$  &  1.146  &  1.00 \\
    H2$\textsubscript{ammon}$-H6$\textsubscript{chain}$  &  1.155  &  1.22 \\
    \hline
    $C2$-NH$_{24}$ (300~GPa)   &  &  \\
    \hline
    H9$\textsubscript{mol}$-H9$^{\prime}\textsubscript{mol}$   &  0.733  &  6.86 \\
    H10$\textsubscript{mol}$-H11$\textsubscript{mol}$     &  0.744  &  6.38 \\
    H7$\textsubscript{mol}$-H8$\textsubscript{mol}$    &  0.764  &  5.94 \\
    H5$\textsubscript{chain}$-H6$\textsubscript{chain}$    &  0.797  &  5.14 \\
    H2$\textsubscript{chain}$-H3$\textsubscript{chain}$    &  0.842  &  4.60 \\
    H2$\textsubscript{chain}$-H4$\textsubscript{chain}$    &  0.961  &  2.87 \\
    H3$\textsubscript{chain}$-H4$\textsubscript{chain}$    &  0.969  &  2.83 \\
    H4$\textsubscript{chain}$-H5$\textsubscript{chain}$    &  0.975  &  2.68 \\
    H6$\textsubscript{chain}$-H8$\textsubscript{mol}$    &  1.029  &  1.96 \\
    H1$\textsubscript{ammon}$-H5$\textsubscript{chain}$    &  1.114  &  1.13 \\
    H3$\textsubscript{chain}$-H7$\textsubscript{mol}$    &  1.148  &  1.14 \\
    H1$\textsubscript{ammon}$-H6$\textsubscript{chain}$  &  1.154  &  0.93 \\
    \hline\hline
    \end{tabular}
    \caption{H-H distances in the NH$_n$ ($n=10, 11, 24$) phases, along with the corresponding crystal orbital Hamiltonian populations integrated to the Fermi level (-ICOHP) at 300~GPa. Values are also provided for solid H$_2$ at 1~atm ($P6_3/m$) and 300~GPa ($Cmca$). The hydrogen atom numbering scheme is provided in Figures \ref{fig:pnma}, \ref{fig:cmc21} and \ref{fig:c2}. H$_\text{mol}$, H$_\text{chain}$ and H$_\text{ammon}$ label hydrogens comprising molecular H$_2$, 1D-chain, or ammonium cation units, respectively.}
    \label{tab:bond-lenght-icohp}
\end{table}

Because the NH$_4$ molecule can be thought of as a superalkali metal atom whose properties resemble those of K or Rb, the formula of $Pnma$-NH$_{10}$ can be written as (NH$_4$)H$_6$, in further analogy with the previously predicted RbH$_6$ and SrH$_6$ phases. However, in stark contrast to the bonding that is observed within these ``simple'' hexahydrides, weak covalent bonds are formed between the hydrogen atoms in the NH$_4^+$ superalkali ion, and the hydrogens within the hydrogenic chains. As Table~\ref{tab:bond-lenght-icohp} reveals these interactions (H(1)-H(5), H(2)-H(4) and H(3)-H(4)), though small, are not negligible, and are of similar magnitude as those between the second nearest neighbor hydrogens within $Cmca$-H$_2$ at 300~GPa.

RbH$_6$, SrH$_6$ and $Pnma$-NH$_{10}$ are all good metals with a high density of states (DOS) at $E\textsubscript{F}$. Their metallicity is in part due to the formation of 1D hydrogenic chains, whose antibonding levels are partially filled via donation from the electropositive alkali or superalkali atom. However, the weak interactions formed between the hydrogen atoms within NH$_4^+$ and those comprising the $^1_\infty$[H$_2$$\cdots$H$^-$] chains in $Pnma$-NH$_{10}$ ensures that character from hydrogens in both the anionic and cationic sublattices are found at $E\textsubscript{F}$, as is evident in the projected DOS (Figure~\ref{fig:pnma}C). Moreover, plots of the -ICOHP averaged over the N-H bonds within NH$_4^+$ (blue curve) and over all H-H pairs within the 1D zig-zag chains (red curve) show that all of the bonding levels are filled, and the antibonding levels are partially occupied at $E\textsubscript{F}$. The bonding states between the hydrogens comprising NH$_4^+$ and the hydrogen chain (pink curve), however, are not completely filled. These weak but numerous covalent interactions may prevent the melting of the 1D chain. Our molecular dynamics (MD) simulations confirmed the thermal stability of $Pnma$-NH$_{10}$ at 150~K and 300~GPa. (Figure~S6A) Although at 200~K some protons started to diffuse to nearby empty lattice sites, the chains do not melt until 450~K, greatly exceeding the $T_\mathrm{c}$ calculated for this compound, whereas the $^1_\infty$[H$_3^-$] chains within RbH$_6$ exhibited liquid-like behavior at comparable temperatures~\cite{Hooper2012_ChemAEurJ}.

\begin{figure*}[!hbtp]
    \centering
    \includegraphics[width=1.0\textwidth]{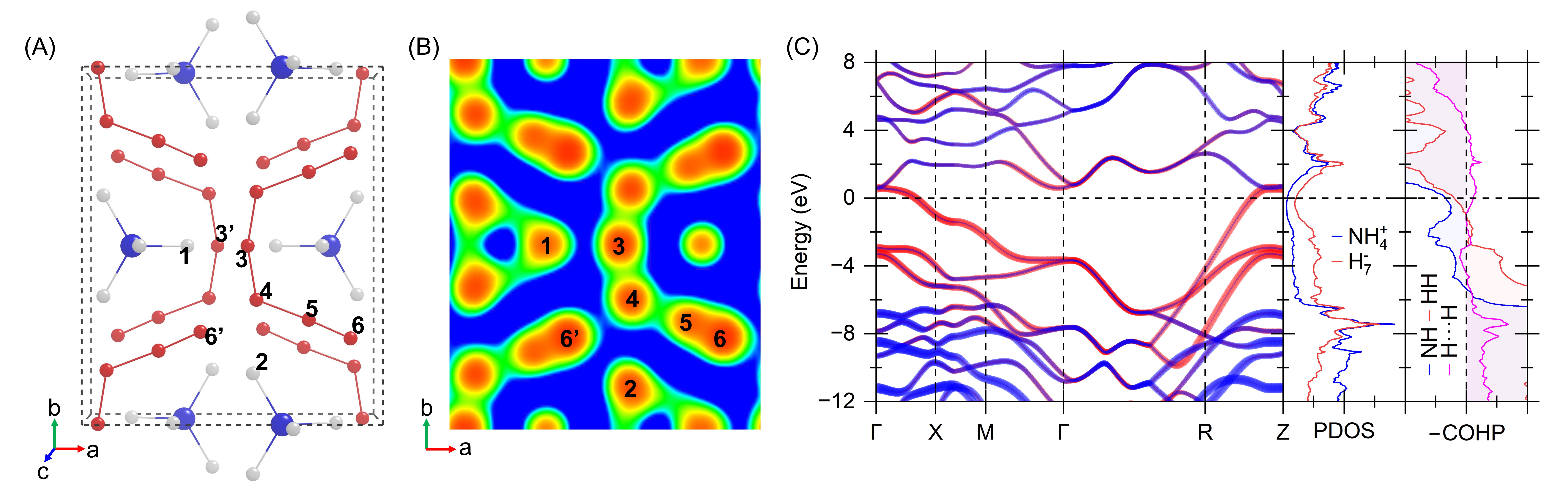}
    \caption{$Cmc2_1$-NH$_{11}$ at 300~GPa: (A) Standard conventional unit cell. (B) ELF profile in a plane that cuts through the H$_7^-$ molecular fragments, colored using the same scheme as in Figure~\ref{fig:pnma}. (C) Atom projected band structure and density of states showing contributions from atoms within the NH$_4^+$ units (blue) and atoms comprising the H$_7^-$ motifs (red). $-$COHP averaged over N-H bonds within NH$_4^+$ (blue), H-H bonds within the H$_7^-$ units (red), and weak  H1-H3, H4-H6$^\prime$, H1-H3$^\prime$, and H2-H6 interactions as in Table~\ref{tab:bond-lenght-icohp} (pink).}
    \label{fig:cmc21}
\end{figure*}

Another CSP-found phase that satisfied our criteria for further analysis, $Cmc2_1$-NH$_{11}$, was computed to be slightly preferred over its neutral molecular constituents ($\Delta H_\text{F}=-7.8$~meV/atom at 300~GPa, with dispersion and ZPE corrections), and it was dynamically stable above 280~GPa (Figure~S4B). The NH$_4^+$ superalkali cation was also a key constituent of this phase, with neighboring cations stacked along the $c$-axis (Figure~\ref{fig:cmc21}A). Along the $a$ and $b$ axes the NH$_4^+$  units were separated from each other via H$_7^-$ molecular fragments, which can be readily identified in a 2D plot of the ELF that passes through them  (Figure~\ref{fig:cmc21}B). This non-planar H$_7^-$ $C_{s}$ symmetry anion contained two tightly bonded H$_2$ units (H(5)-H(6)) with Bader charges of -0.10.  The H(5)-H(4) and H(4)-H(3) distances were $\sim$0.16~\AA{} longer than in these strongly bonded dihydrogen units resulting in interactions that were 2.5~eV weaker, on average (Table \ref{tab:bond-lenght-icohp}). The Bader charge of H(4) was +0.08, and that of H(3) was -0.12. This charge distribution differs from the one computed for an H$_7^-$ anion in the gas phase because of the  numerous bonds present between the hydrogens comprising the superalkali cation and the C-shaped H$_7^-$ anion. However, within $Cmc2_1$-NH$_{11}$ these interactions are even weaker than in $Pnma$-NH$_{10}$, as evidenced by the longer bond distances, smaller -ICOHPs, as well as a decreased NH$_4^+$ projected DOS and -COHP between the respective hydrogen atoms at $E\textsubscript{F}$ (Figure~\ref{fig:cmc21}C). MD simulations revealed that $Cmc2_1$-NH$_{11}$ is thermally stable at 300~GPa and 100~K, but by 150~K the NH$_4^+$ molecules rotate as the N-H bonds vibrate (Figure~S6B).

Finally, we describe the structural peculiarities of the most dilute hydride of ammonia that was analyzed, $C2$-NH$_{24}$. Prior high pressure experiments have synthesized compounds with high weight percent H$_2$ content including HI(H$_2$)$_{13}$ (17.7\%)~\cite{Binns2018_PhysRevB} and (H$_2$)$_8$CH$_4$ (33.4\%)~\cite{Mao2004_PNAS}. The phase predicted here contains an impressive 52.6 or 55.3 weight \% H$_2$ (when NH$_4$ or NH$_3$ are treated as indivisible entities, respectively). It is dynamically stable from 200-380~GPa (Figure~S4C), with a $\Delta H_\text{F}$ of 3.8 and 5.3~meV/atom at 200 and 300~GPa, respectively, when ZPE and dispersion corrections are included. MD simulations showed that $C2$-NH$_{24}$ is stable up to 150~K at 300~GPa (Figure~S6C).

\begin{figure*}[!htbp]
    \centering
    \includegraphics[width=1.0\textwidth]{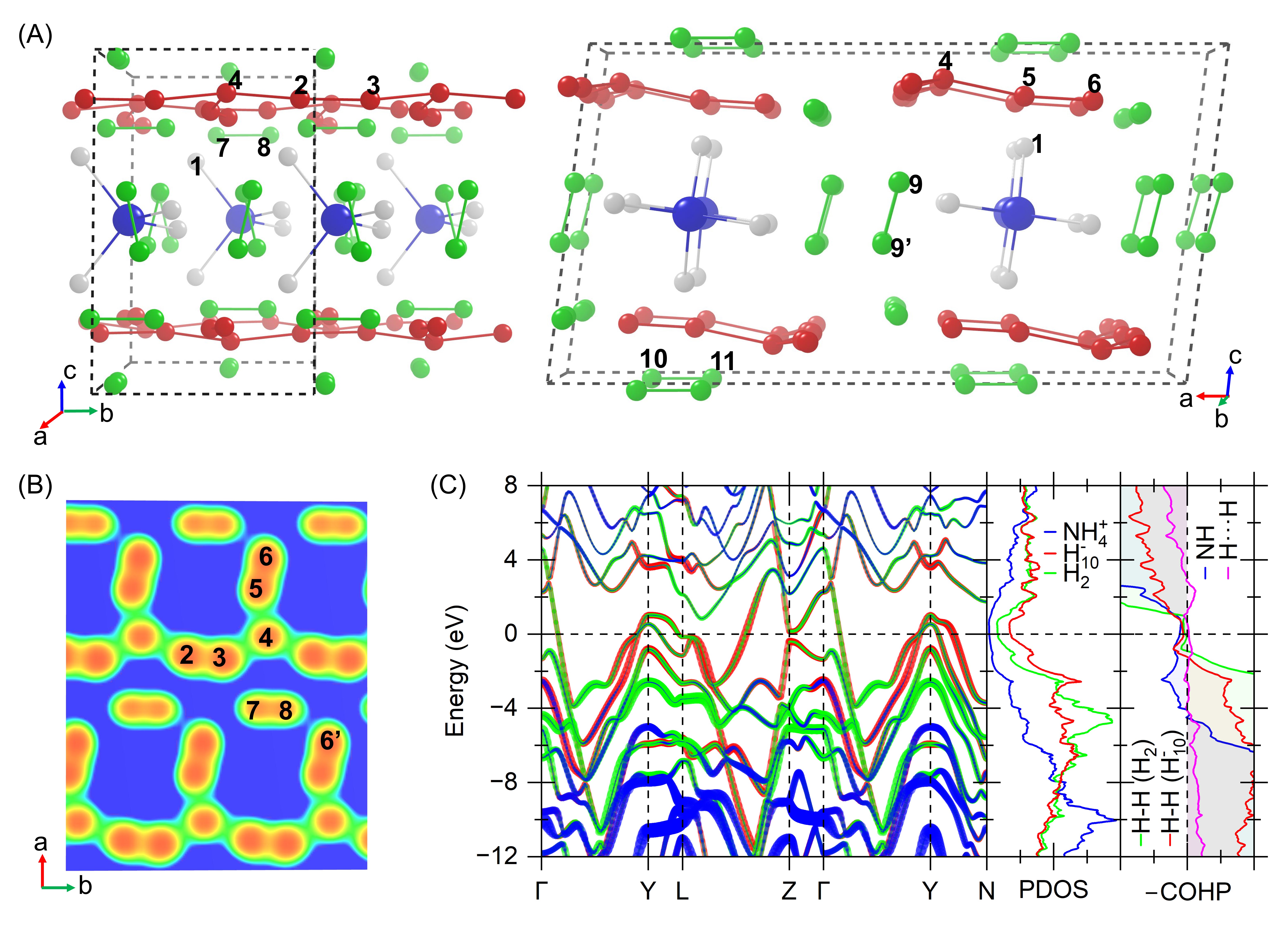}
    \caption{$C2$-NH$_{24}$ at 300~GPa: (A) Standard conventional unit cell from (left) front view and (right) side view. (B) ELF profile in a plane that cuts through the 1D chain, colored using the same scheme as in Figure~\ref{fig:pnma}. (C) Atom projected band structure and density of states showing contributions from atoms within the NH$_4^+$ units (blue), atoms within the H$_2$ units (green), and atoms within the 1D chains (red). $-$COHP averaged over N-H bonds within NH$_4^+$ (blue), H-H bonds within the H$_2$ units (green), H-H bonds within the 1D chain (red), and between the H atoms within NH$_4^+$ or H$_2$ and those in the chain (H6-H8, H1-H5, H3-H7, and H1-H6 as in Table~\ref{tab:bond-lenght-icohp}, pink).}
    \label{fig:c2}
\end{figure*}

The standard conventional cell of $C2$-NH$_{24}$ (Figure~\ref{fig:c2}A) contains two NH$_4^+$ molecules stacked parallel to the $b$-axis. In the $ab$ plane these rows of ammonium cations are separated from one another by two 1D polymeric hydrogen motifs (H(2) to H(6)) that themselves are separated by H$_2$ molecules (H(10)-H(11)), and a further set of H$_2$ molecules separate the superalkali cations in the $bc$ plane (H(9)-H(9$^\prime$)).  Therefore, another way to write the formula of this phase is NH$_4$(H$_{10}$)(H$_2$)$_{5}$. Turning to the 1D chain, whose ELF profile is plotted in Figure~\ref{fig:c2}B, reveals pairs of hydrogens that form shorter stronger bonds (H(5)-H(6) and H(2)-H(3)) surrounding a nearly neutral lone hydrogen atom (H(4)).  The bond distances and strengths of the ensuing interactions are provided in Table \ref{tab:bond-lenght-icohp}. Similar to the previously discussed phases, weak but numerous interactions between the hydrogen atoms within NH$_4^+$ and those within the anionic sublattice are found. However, $C2$-NH$_{24}$ is predicted to be dynamically stable within the harmonic approximation to pressures as low as 200~GPa, whereas $Pnma$-NH$_{10}$ required at least 280~GPa to prevent the onset of a dynamic instability. The lowering of the stabilization pressure might be attributed to the dilution of the negative charge on the extended hydrogenic framework from H$_6^-$ in $Pnma$-NH$_{10}$ to H$_{10}^-$ in $C2$-NH$_{24}$. Thus, we postulate that phases containing an even larger weight percent hydrogen, with extensive polymeric hydrogen chains, could be (meta)stable at lower pressures, in particular when anharmonic fluctuations, likely to be important for these hydrogen atoms, are considered. 

As the band structure and DOS plots at 300~GPa show (Figure \ref{fig:c2}A), $C2$-NH$_{24}$ is metallic in the whole pressure range of its stability. Here, a \emph{little} bit of nitrogen has really done much for hydrogen!\cite{Zurek:2009c}
The formal charge on the anionic lattice in $C2$-NH$_{24}$ (assuming full transfer of the unpaired electron in NH$_4$ to the hydrogenic lattice) is H$_{20}^-$ as compared to H$_6^-$ for NH$_{10}$. As a result, there are fewer antibonding states filled within NH$_{24}$, which will decrease further in the limit of infinite dilution. 

Let us now explore the propensity for superconductivity in these superhydride phases and identify the vibrations that furnish the greatest contribution to their predicted $T\textsubscript{c}$s. To do so, we calculated the Eliashberg spectral function, $\alpha^2\textrm{F}(\omega)$, and from it we obtained the electron-phonon-coupling (EPC) parameter, $\lambda$, as well as the logarithmic average of the phonon frequencies, $\omega\textsubscript{log}$. The combined nuclear quantum and anharmonic effects, which could be significant in these systems, were neglected, the Coulomb parameter, $\mu^*$, was set to 0.1, and $T\textsubscript{c}$ was estimated via the Allen-Dynes modified McMillan (ADM) equation (including the strong coupling and shape dependence)~\cite{Allen1975_PhysRevB}, and within isotropic Eliashberg theory\cite{Eliashberg1960_SPJ}.

\begin{figure*}
    \centering
    \includegraphics[width=1.0\textwidth]{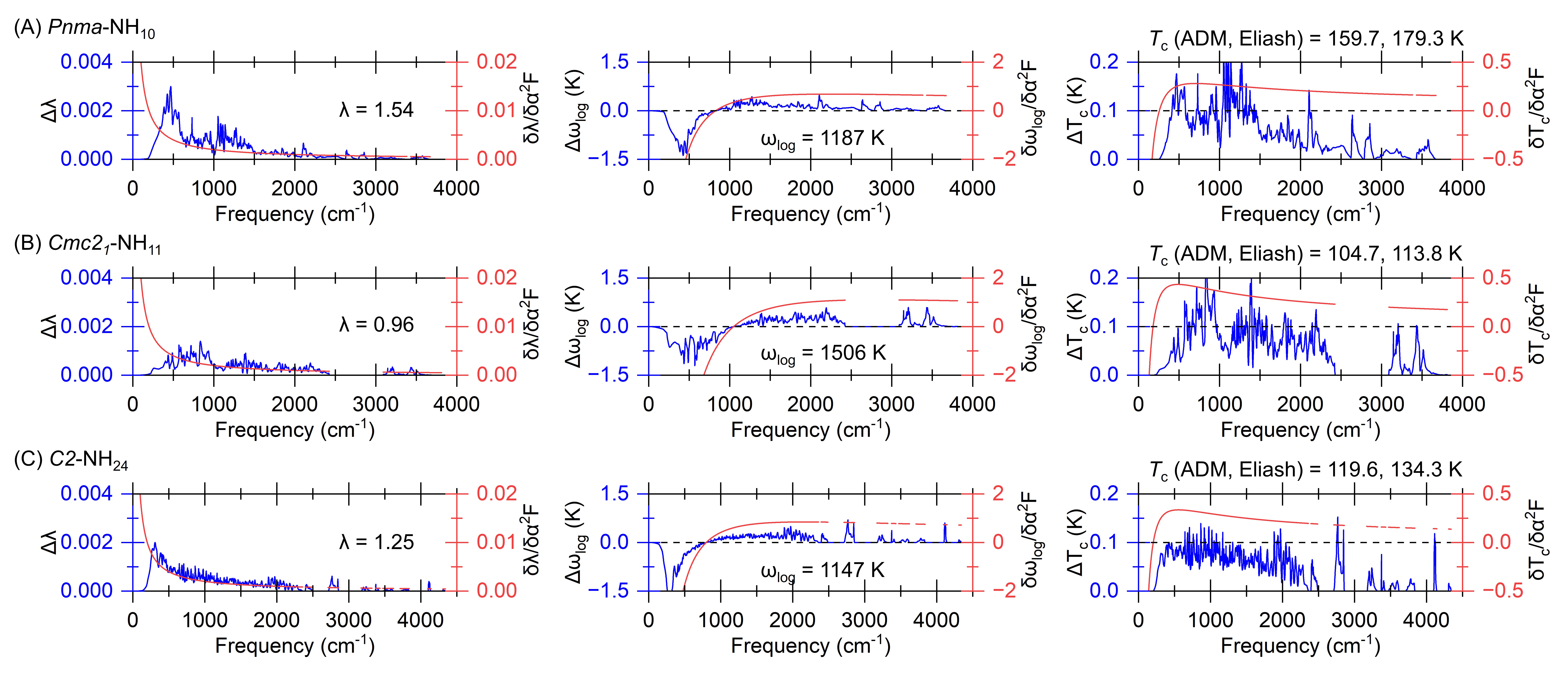}
    \caption{Superconducting properties of (a) $Pnma$-NH$_{10}$, (b) $Cmc2_1$-NH$_{11}$, and (c) $C2$-NH$_{24}$ at 300~GPa. Contribution of the vibrational frequencies to the electron-phonon-coupling ($\Delta\lambda$), logarithmic average frequency ($\Delta\omega_\textrm{log}$) and superconducting critical temperature ($\Delta T_\textrm{c})$ as a function of frequency (blue curves). The functional derivatives with respect to the Eliashberg spectral function, $\delta\lambda/\delta\alpha^2\textrm{F}$, $\delta\omega\textsubscript{log}/\delta\alpha^2\textrm{F}$ and $\delta T\textsubscript{c}/\delta\alpha^2\textrm{F}$, are also plotted (red curves). The $T_\text{c}$ predicted with the Allen-Dynes modified McMillan (ADM) and Eliashberg equations and $\mu^*=0.1$ is provided. }
    \label{epc}
\end{figure*}

One way to identify the phonons most important for increasing the $T\textsubscript{c}$ in a conventional superconductor is by calculating its functional derivative with respect to the Eliashberg spectral function, $\delta T\textsubscript{c}/\delta\alpha^2\textrm{F}$ \cite{Bergmann1973_ZPhys,Allen1975_PhysRevB}. Such an analysis has previously been applied to compressed hydrides, revealing the important role of bending vibrations at low pressures, while at high pressures mixed bending and stretching vibrations spanning a wide range of frequencies contributed to the EPC~\cite{Tanaka2017_PhysRevB}. Using the ADM, we calculated the functional derivative not only of $T\textsubscript{c}$, but also of $\lambda$ and $\omega\textsubscript{log}$ numerically using a frequency interval of 1~cm$^{-1}$. The results for the three studied superhydrides at 300~GPa are plotted in Figure~\ref{epc} (red curves). An additional analysis was performed where $\alpha^2\textrm{F}(\omega)$ was set to zero in a frequency range between $\omega$ to $\omega+\Delta\omega$, and the superconducting properties were recomputed. The change induced by this perturbation is plotted as $\Delta\lambda$, $\Delta\omega\textsubscript{log}$ and $\Delta T\textsubscript{c}$ in the same figures (blue curves). 

The plot of $\Delta\lambda$ vs.\ $\omega$ for $Pnma$-NH$_{10}$ showed that modes whose frequencies were below 490~cm$^{-1}$ contributed most to $\lambda$. Further analysis revealed that the mode with the strongest EPC, \emph{i.e.}~largest $\lambda_{\mathbf{q}\nu}$, is found at the first phonon branch at the $X$ point with a frequency of 180~cm$^{-1}$. Visualization of this mode showed that it is an H(4)-H(5)-H(6) stretch coupled with an NH$_4^+$ rotation. The former, which corresponds to vibrations of the 1D chain, could result in the formation of H$_2$/H$^-$ units with the concomitant opening of a gap near $E\textsubscript{F}$ along the $\Gamma$-$Z$ high symmetry line (Figure~\ref{fig:pnma}). The  bands that cross $E\textsubscript{F}$ along $\Gamma$-$X$-$S$-$Y$-$\Gamma$ exhibit character from both the anionic and cationic sublattices, as evident in a plot of the partial charge density calculated within $\pm$0.15~eV of the Fermi level. This plot showed the H$_6^-$ chains enveloped by tubes connected via charge density arising from the ammonium cations, by way of a  rhombus formed from H(2), H(3), and two H(4)s from two neighbouring H$_6^-$ chains (Figure~S7). Rotation of the NH$_4^+$ elongates the distances within the H$_4$ rhombus, and perturbs the numerous weak interactions between the hydrogens comprising the anionic and cationic lattices, resulting in the EPC.

Visualization of a subset of (randomly chosen) phonon branches shows they resemble the mode with the largest $\lambda_{\mathbf{q}\nu}$, therefore one might expect that the phonon linewidth ($\gamma_{\mathbf{q}\nu}$) for most of the phonon modes would be similar, as confirmed by a plot of the phonon band structure decorated by the $\gamma_{\mathbf{q}\nu}$ (Figure~S8). Despite their large and positive contribution towards the overall $\lambda$, the low frequency modes below $\sim$500~cm$^{-1}$ decrease the overall $\omega\textsubscript{log}$ (and are thus shown with a negative sign in the middle panel in Figure \ref{epc}A), and their overall contribution to  $T\textsubscript{c}$ (right panel) is small. In fact setting the $\alpha^2\textrm{F}(\omega)$ values below 490~cm$^{-1}$ to zero decreases the $T\textsubscript{c}$ by only 16 (24)~K in the ADM (Eliashberg) approximation. A similar analysis for the phonon branches above 2330~cm$^{-1}$, which are linear combinations of N-H and H-H stretching modes, yields a drop of the  ADM (Eliashberg) $T\textsubscript{c}$ by only 23 (28)~K. Therefore, similar to Ref.\ \cite{Tanaka2017_PhysRevB}, we find that the intermediate frequency regime from 490 to 2300~cm$^{-1}$ contributes most to the superconductivity. The maximum of the functional derivative, $\delta T\textsubscript{c}/\delta\alpha^2\textrm{F}$, which defines the optimal frequency ($\omega\textsubscript{opt}$), can be used to estimate the $T\textsubscript{c}$ as $\omega\textsubscript{opt}\sim7k\textsubscript{B}T\textsubscript{c}$~\cite{Bergmann1973_ZPhys}, where $k\textsubscript{B}$ is the Boltzmann constant, yielding $\sim$169~K, which is close to our Allen-Dynes value of 160~K, and slightly smaller than our Eliashberg value of 180~K.

$Cmc2_1$-NH$_{11}$ has a smaller DOS at $E_\textrm{F}$ (0.069 states/Ry/\r{A}$^3$) as compared to $Pnma$-NH$_{10}$ (0.108 states/Ry/\r{A}$^3$), and a plot of the band structure decorated by the  $\gamma_{\mathbf{q}\nu}$ (Figure~S9) illustrates that only a few of the phonon modes have large linewidths, in contrast to the decahydride. Indeed, Figure~\ref{epc}B illustrates that $Cmc2_1$-NH$_{11}$ possesses fewer low frequency modes with a high EPC, resulting in a smaller total $\lambda$, but a larger $\omega\textsubscript{log}$. The ADM and Eliashberg predicted $T\textsubscript{c}$ values are similar (105 vs.\ 114~K). The lower critical temperature of NH$_{11}$ (as compared to NH$_{10}$) is not surprising as it possesses a molecular hydrogenic sublattice, in agreement with the trends in $T\textsubscript{c}$ that have been observed for other high pressure polyhydrides with electropositive elements~\cite{Zurek2019_JChemPhys}. 
Finally, $C2$-NH$_{24}$ with both molecular and 1D-periodic hydrogenic motifs falls somewhere between NH$_{10}$ and NH$_{11}$ in terms of the three functional derivatives, the way in which certain frequencies contribute to the integrated superconducting parameters, and the DOS at $E_\textrm{F}$ (0.064 states/Ry/\r{A}$^3$). Therefore, it should not be a surprise that its estimated $T\textsubscript{c}$ is also intermediate, with a 120~K ADM (134~K Eliashberg) value.   

In conclusion, evolutionary structure searches predicted a number of metallic, high-symmetry NH$_n$ phases that were metastable under pressure. These compounds, containing ammonium cations and a wide variety of hydrogenic lattices, including 1D hydrogenic chains, dihydrogen, and more complex hydrogenic molecular units, were predicted to be conventional superconductors. Because the quasi-spherical NH$_4^+$ molecule can be thought of as a superalkali atom, with a radius and ionization potential resembling that of K or Rb, these compounds can be seen as extensions of the alkali metal polyhydrides. The  studied NH$_{10}$, NH$_{11}$ and NH$_{24}$ compounds, whose formulae could be written as (NH$_4$)(H$_6$), (NH$_4$)(H$_7$) and NH$_4$(H$_2$)$_5$H$_{10}$,  afforded weak bonds between the hydrogens comprising the NH$_4^+$ cation and those in the anionic lattice, with H-H distances of $\sim$1-1.2~\AA{}. Analysis of the electron phonon coupling mechanism showed that rotations of the NH$_4^+$ molecules modify these weak interactions, thereby perturbing the hydrogen atoms in the anionic sublattice and contributing to the microscopic mechanism of superconductivity. The predicted (Eliashberg) $T_\textrm{c}$s, of 179, 114, and 134~K for NH$_{10}$, NH$_{11}$ and NH$_{24}$, respectively, at 300~GPa, are comprable to some of the higher $T_\textrm{c}$s computed for the alkali polyhydrides~\cite{Shipley2021_PhysRevB}. Our study suggests that molecular cations can serve the role of the electropositive element in superconducting hydrides, including ones that may be constituents of the interiors of gas giant planets.

\begin{acknowledgement}
Funding for this research is provided by the Center for Matter at Atomic Pressures (CMAP), a National Science Foundation (NSF) Physics Frontier Center, under Award PHY-2020249, and the the NSF award DMR-2136038. Calculations were performed at the Center for Computational Research at SUNY Buffalo (http://hdl.handle.net/.10477/79221) and National Energy Research Scientific Computing Center (NERSC) and Livermore Computing.
\end{acknowledgement}

\begin{suppinfo}
The Supporting Information is available free of charge on the ACS Publication website. It includes full computational details, 2D convex hulls, molecular orbitals of the NH$_4$ radical, phonon band structures and density of states, electronic structure analysis, trajectories of molecular dynamics runs, projected phonon linewidths and Eliashberg spectral functions, and structural parameters. 
\end{suppinfo}

\bibliography{reference}

\end{document}